\pdfoutput=1

\documentclass[10pt,a4paper]{llncs}

\usepackage{amsmath}
\usepackage{graphicx}
\usepackage{booktabs}
\usepackage{amssymb}
\usepackage{latexsym}	
\usepackage{array}		
\usepackage{multirow}
\usepackage{subfigure}
\usepackage{algorithm}
\usepackage{algpseudocode}
\usepackage{threeparttable}
\usepackage{paralist}   
\usepackage{xspace}
\usepackage{color}
\usepackage{xcolor}
\usepackage{adjustbox}
\usepackage{balance}
\usepackage{wasysym}
\usepackage{rotating}   
\usepackage{listings}
\usepackage{flushend}
\usepackage{caption}    

\usepackage{hyperref}
\usepackage{url}            

\newcommand{\red}[1]{\textcolor[rgb]{1.00,0.00,0.00}{#1}}

\definecolor{mygreen}{rgb}{0,0.6,0}
\definecolor{mygray}{rgb}{0.5,0.5,0.5}
\definecolor{mymauve}{rgb}{0.58,0,0.82}
\definecolor{darkblue}{rgb}{0.0,0.0,0.6}
\definecolor{maroon}{RGB}{102, 0, 0}
\definecolor{Maroon}{cmyk}{0,0.87,0.68,0.32}
\definecolor{darkred}{RGB}{139, 0, 0}

\lstdefinelanguage{XML}
{
  basicstyle=\ttfamily\small,   
  morestring=[b]",
  moredelim=[s][\color{darkblue}]{<}{\ },
  moredelim=[s][\color{darkblue}]{</}{>},
  moredelim=[l][\color{darkblue}]{/>},
  moredelim=[l][\color{darkblue}]{>},
  morecomment=[s]{<?}{?>},
  morecomment=[s]{<!--}{-->},
  stringstyle=\color{darkred},
  identifierstyle=\color{mymauve}
}

\newif\ifACM
\ACMfalse 

\ifACM
\newcommand{\myfig}{Figure\xspace}
\else
\newcommand{\myfig}{Fig.\xspace}
\fi

\newcommand{\minSDK}{\texttt{minSdkVersion}\xspace}
\newcommand{\aimSDK}{\texttt{targetSdkVersion}\xspace}
\newcommand{\maxSDK}{\texttt{maxSdkVersion}\xspace}
\newcommand{\DSDK}{\texttt{DSDK}\xspace}
\newcommand{\lagSDK}{\texttt{lagSdkVersion}\xspace}

\newcommand{\minLevel}{\texttt{minLevel}\xspace}
\newcommand{\maxLevel}{\texttt{maxLevel}\xspace}

\newcommand{\aapt}{\texttt{aapt}\xspace}
\newcommand{\apktool}{\texttt{apktool}\xspace}
\newcommand{\dexdump}{\texttt{dexdump}\xspace}

\newcommand{\repeatthanks}{\textsuperscript{\thefootnote}}

\begin{document}
	
	\mainmatter  
	
	\title{Measuring the Declared SDK Versions and Their Consistency with API Calls in Android Apps}
	
	\titlerunning{}
	
    \author{Daoyuan Wu \and Ximing Liu\thanks{These two author names are in alphabetical order.} \and Jiayun Xu\repeatthanks \and David Lo \and Debin Gao}
	\authorrunning{}
	
    \institute{School of Information Systems, Singapore Management University, Singapore.\\
      \email{\{dywu.2015,xmliu.2015,jyxu.2015,davidlo,dbgao\}@smu.edu.sg}
      \\
      \red{This paper will appear in WASA 2017 (\url{http://wasa-conference.org/WASA2017/}).}\\
      \red{It is originally a course project paper done by the first three authors in April 2016.}
    }

\mainmatter
\maketitle

\vspace{-4ex}
\begin{abstract}

Android has been the most popular smartphone system, with multiple platform versions (e.g., KITKAT and Lollipop) active in the market.
To manage the application's compatibility with one or more platform versions, Android allows apps to declare the supported platform SDK versions in their manifest files.
In this paper, we make a first effort to study this modern software mechanism.
Our objective is to measure the current practice of the declared SDK versions (which we term as \DSDK versions afterwards) in real apps, and the consistency between the \DSDK versions and their app API calls.
To this end, we perform a three-dimensional analysis.
First, we parse Android documents to obtain a mapping between each API and their corresponding platform versions.
We then analyze the \DSDK-API consistency for over 24K apps, among which we pre-exclude 1.3K apps that provide different app binaries for different Android versions through Google Play analysis.
Besides shedding light on the current \DSDK practice, our study quantitatively measures the two side effects of inappropriate \DSDK versions: (i) around 1.8K apps have API calls that do not exist in some declared SDK versions, which causes runtime crash bugs on those platform versions; (ii) over 400 apps, due to claiming the outdated targeted \DSDK versions, are potentially exploitable by remote code execution.
These results indicate the importance and difficulty of declaring correct \DSDK, and our work can help developers fulfill this goal.

\end{abstract}

\vspace{-4ex}
\section{Introduction}
\label{sec:intro}
\vspace{-1ex}

Recent years have witnessed the extraordinary success of Android, a smartphone operating system owned by Google.
At the end of 2013, Android became the most sold phone and tablet OS.
As of 2015, Android evolved into the largest installed base of all operating systems.
Along with the fast-evolving Android, its fragmentation problem becomes more and more serious.
Although new devices ship with the recent Android versions, there are still huge amounts of existing devices running old Android versions~\cite{dashboards}.

To better manage the application's compatibility with multiple platform versions, Android allows apps to declare the supported platform SDK versions in their manifest files.
We term these declared SDK versions as \DSDK versions.
The \DSDK mechanism is a modern software mechanism that to the best of our knowledge, few systems are equipped with such mechanism until Android.
Nevertheless, so far the \DSDK receives little attention and few understandings are known about the effectiveness of the \DSDK mechanism.

In this paper, we make a first attempt to systematically study the \DSDK mechanism.
In particular, our objective is to measure the current practice of \DSDK versions in real apps, and the consistency between \DSDK versions and their apps' API calls.
To this end, we perform a three-dimensional analysis that analyzes Google Play, Android documents, and each individual app.
We use a large dataset that contains over 24K apps crawled from Google Play in July 2015.
Our study sheds light on the current \DSDK practice and quantitatively measures the two side effects of inappropriate \DSDK versions.

We summarize the contributions of this paper as follows:
\begin{compactitem}
\item (\textit{New problem}) We study a modern software mechanism, i.e., allowing apps to declare the supported platform SDK versions. In particular, we are the first to measure the declared SDK versions and their consistency with API calls in Android apps.
  
\item (\textit{New understanding}) We give the first demystification of the \DSDK mechanism and its two side effects of inappropriate \DSDK versions. 

\item (\textit{Hybrid approach}) We propose a three-dimensional analysis method that operates at both Google Play, Android document, and Android app levels.
  
\item (\textit{Insightful results}) We have three major findings, including (i) around 17\% apps do not claim the targeted \DSDK versions or declare them wrongly, (ii) around 1.8K apps under-set the minimum \DSDK versions, causing them crash when running on lower Android versions, and (iii) over 400 apps under-claim the targeted \DSDK versions, making them potentially exploitable by remote code execution.

\end{compactitem}


\section{Demystifying The Declared SDK Versions and Their Two Side Effects}
\label{sec:backg}

In this section, we first demystify the declared platform SDK versions in Android apps, and then explain their two side effects if inappropriate \DSDK versions are being used.

\vspace{-1.5ex}
\subsection{Declared SDK Versions in Android Apps}
\label{sec:declared}

\vspace{-2ex}
\begin{lstlisting}[
language=XML,
basicstyle=\ttfamily\small,
label={lst:sdkversion},
caption={\small The syntax for declaring the platform SDK versions in Android apps.}
]
<uses-sdk android:minSdkVersion="integer"
        android:targetSdkVersion="integer"
        android:maxSdkVersion="integer" />
\end{lstlisting}

Listing \ref{lst:sdkversion} illustrates how to declare the supported platform SDK versions in Android apps by defining the \texttt{<uses-sdk>} element in apps' manifest files (i.e., \texttt{AndroidManifest.xml}).
These \DSDK versions are for the runtime Android system to check apps' compatibility, which is different from the compiling-time SDK for compiling source codes.
The value of each \DSDK version is an integer, which represents the API level of the corresponding SDK.
For example, if a developer wants to declare the SDK version 5.0, he/she sets its value as 21 (the API level of Android 5.0 is 21).
Since each API level has a precise mapping of the corresponding SDK version~\cite{AndroidVersion}, we do not use another term, \textit{declared API level}, to represent the same meaning of \DSDK throughout this paper.

We explain the three \DSDK attributes as follows: 
\begin{itemize}
  \item The \minSDK integer specifies the minimum platform API level required for the app to run. The Android system refuses to install an app if its \minSDK value is greater than the system's API level. Note that if an app does not declare this attribute, the system by default assigns the value of ``1'', which means that the app can be installed in all versions of Android.

  \item The \aimSDK integer designates the platform API level that the app targets at. An important \textit{implication} of this attribute is that Android adopts the back-compatible API behaviors of the declared target SDK version, even when an app is running on a higher version of the Android platform. Android makes such compromised design because it aims to guarantee the same app behaviors as developers expect, even when apps run on newer platforms. It is worth noting that if this attribute is not set, the default value equals to the value of \minSDK.

  \item The \maxSDK integer specifies the maximum platform API level on which an app can run. However, this attribute is \textit{not} recommended and already \textit{deprecated} since Android 2.1 (API level 7). That said, modern Android no longer checks or enforces this attribute during the app installation or re-validation. The only effect is that Google Play continues to use this attribute as a filter when it presents users a list of applications available for download. Not that if this attribute is not set, it implies no any restriction on the maximum platform API level.
\end{itemize}


\subsection{Two Side Effects of Inappropriate DSDK Versions}
\label{sec:sideeffect}

\begin{figure}[t!]
\vspace{-4ex}
\begin{adjustbox}{center}
\includegraphics[width=0.6\textwidth]{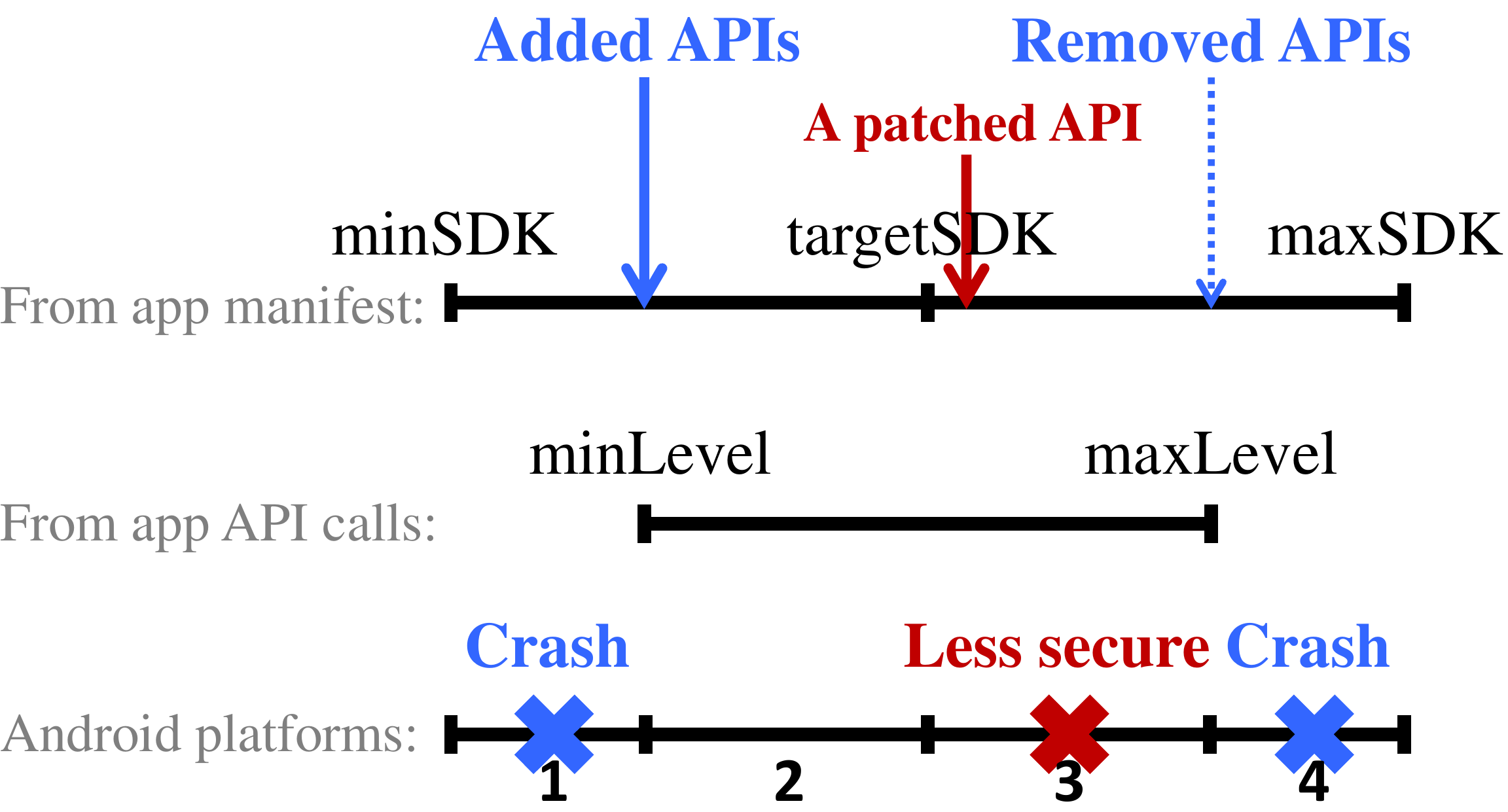}
\end{adjustbox}
\vspace{-4ex}
\caption{Illustrating the two side effects of inappropriate \DSDK versions.}
\vspace{-4ex}
\label{fig:demystify}
\end{figure}

\myfig \ref{fig:demystify} illustrates the two side effects of inappropriate \DSDK versions.
We first explain the symbols used in this figure, and then describe the two side effects in the subsequent paragraphs. 
As shown in \myfig~\ref{fig:demystify}, we can obtain $minSDK$, $targetSDK$, and $maxSDK$ from an app manifest file.
Based on the API calls of an app, we can calculate the minimum and maximum API levels it requires, i.e., $minLevel$ and $maxLevel$. 
Eventually, the app will be deployed to a range of Android platforms between $minSDK$ and $maxSDK$.

\noindent
\textbf{Side Effect I: Causing Runtime Crash Bugs}
The blue part of \myfig \ref{fig:demystify} shows two scenarios in which inappropriate \DSDK versions can cause app crash.
The first scenario is $minLevel > minSDK$, which means a new API is introduced after the $minSDK$.
Consequently, when an app runs on the Android platforms between $minSDK$ and $minLevel$ (marked as the block 1 in \myfig \ref{fig:demystify}), it will crash.
We verified this case by using the \texttt{VpnService.Builder.addDisallowedApplication()} API, which was introduced at Android 5.0 at the API level 21.
We called this API at the MopEye app \cite{MopEyePoster15} and ran MopEye on an Android 4.4 device.
When the app executed the \texttt{addDisallowedApplication()} API, it crashed with the \texttt{java.lang.NoSuchMethodError} exception.

The second crash scenario is $maxSDK > maxLevel$, which means an old API is removed at the $maxLevel$.
Similar to the first scenario, the app will crash when it runs on the Android platforms between $maxLevel$ and $maxSDK$.


\noindent
\textbf{Side Effect II: Making Apps Less Secure}
The red part of \myfig \ref{fig:demystify} shows the scenario in which inappropriate \DSDK versions cause apps fail to be patched that they originally should be able to.
Suppose an app calls an API (e.g., \texttt{addJavascriptInterface()} \cite{addJavascriptInterfaceSaga}) that is vulnerable before the $targetSDK$.
However, if the \texttt{targetSdkVersion} of the app is lower than the patched API level, Android will still take the compatibility behaviors, i.e., the non-patched API behavior in this case, even when the app runs on the patched platforms (between $targetSDK$ and $maxLevel$).
Some such vulnerable app examples are available in \url{https://sites.google.com/site/androidrce/}.

\section{Methodology}
\label{sec:method}

In this section, we present an overview of our methodology and its three major components.

\subsection{Overview}
\label{sec:overview}

\begin{figure}[t!]
\begin{adjustbox}{center}
    \begin{minipage}{0.5\textwidth}
        \centering
        \includegraphics[width=1\textwidth]{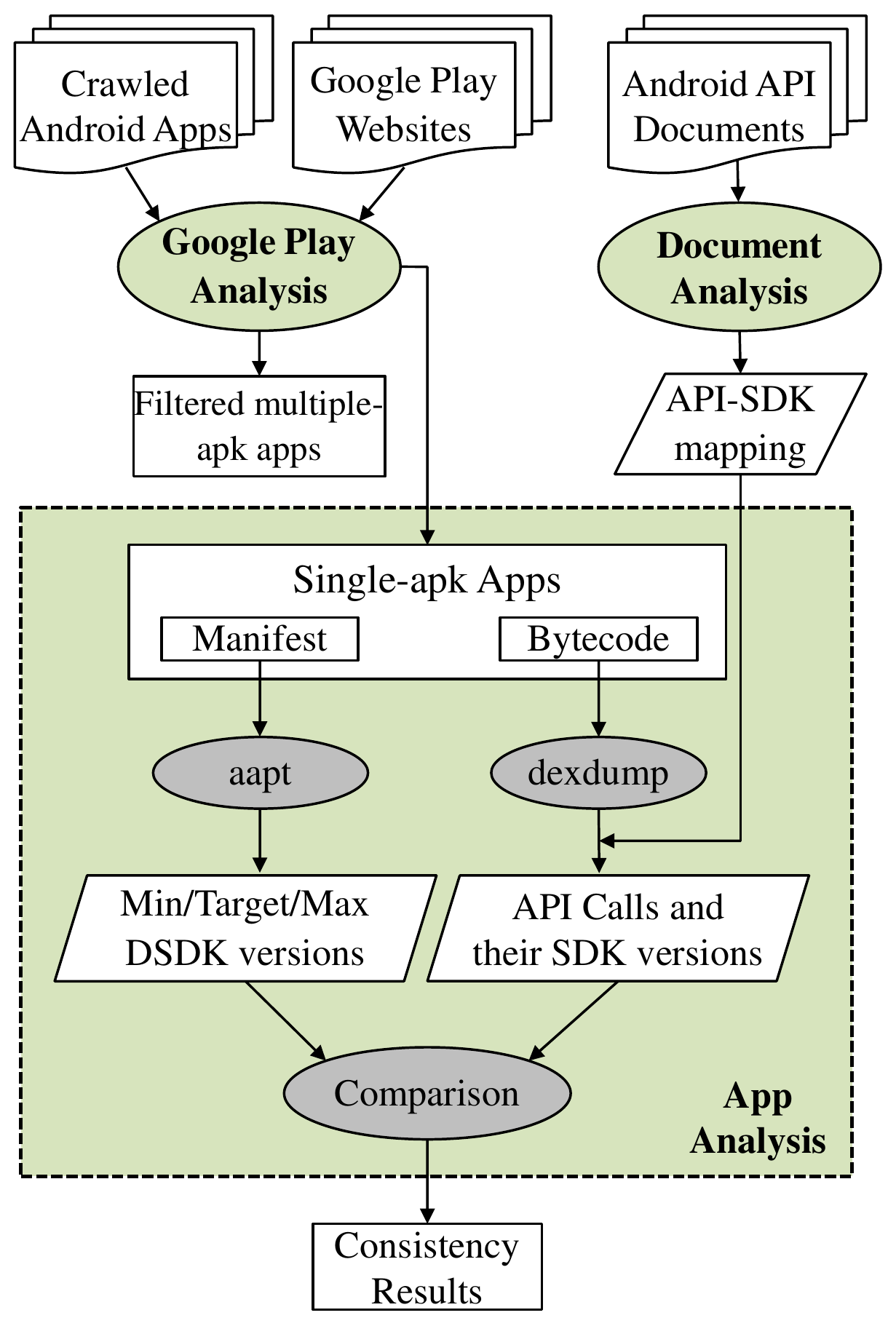}
        \caption{The overview of our methodology.}
        \label{fig:overview}
    \end{minipage}
    \hspace{0ex}
    \begin{minipage}{0.5\textwidth}
        \centering
        \includegraphics[width=1\textwidth]{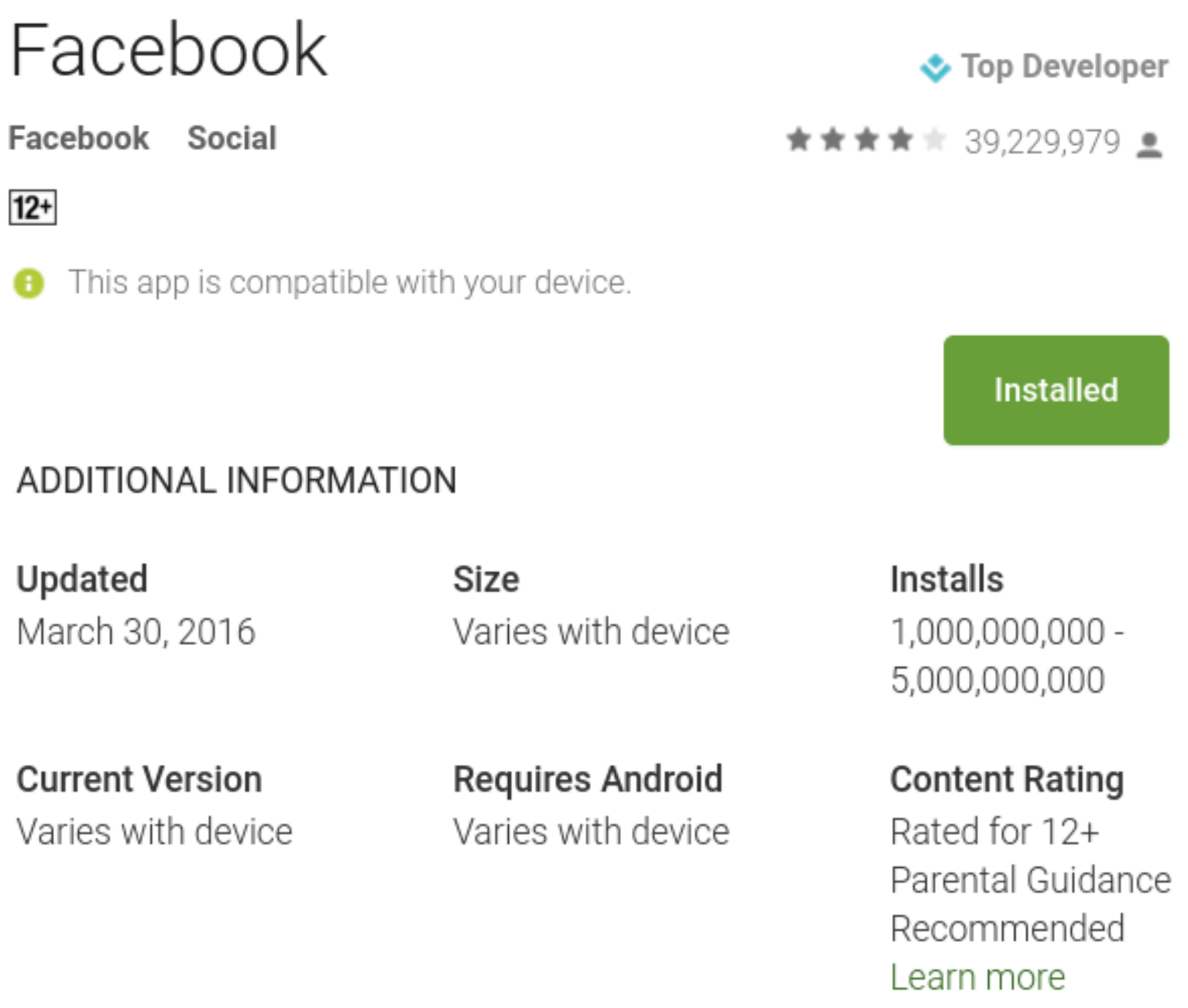}
        \caption{The Facebook app's Google Play page (with irrelevant contents removed).}
        \label{fig:playPage}
        \vspace{5ex}
        \centering
        \captionof{table}{The dataset of our study.}
        \label{tab:dataset}
\begin{tabular}{ c | c | c }
\hline
 & \# & Note \tabularnewline
\hline
\hline
All crawled apps    & 24,426    & The initial dataset \tabularnewline
\hline
Multiple-apk apps   & 1,301     & Filtered apps\tabularnewline
\hline
Single-apk apps     & 23,125    & The final dataset \tabularnewline
\hline
\end{tabular}
\label{tab:dataset}
    \end{minipage}
\end{adjustbox}
\end{figure}

\myfig \ref{fig:overview} illustrates the overall design of our method.
It performs the analysis at three levels.
First, we crawl and analyze each app's Google Play page to filter \textit{multiple-apk} apps that provide different app binaries (i.e., \textit{apks}) for different Android platforms.  
Since each apk of these apps is tailored for a particular Android version, its declared platform SDK version is no longer important.
We therefore exclude these multiple-apk apps for further analysis.
Second, we parse Android API documents to build a complete mapping between each API and their corresponding platform versions.
We call this mapping the \textit{API-SDK mapping}.

In the final app analysis phase, we first extract apps' declared SDK versions and API calls, then leverage the existing API-SDK mapping to infer the range of SDK versions from API calls, and finally compare these two SDK versions (i.e., the declared SDK versions and the SDK versions inferred from API calls). 
The output is the (in)consistency results between declared SDK versions and API calls, which can be further leveraged to detect bugs and vulnerabilities.

\subsection{Google Play Analysis}
\label{sec:PlayAnalysis}

\noindent
\textbf{Design and Implementation}
The main objective of running Google Play analysis is to filter multiple-apk apps.
We explain this step using a representative Google Play page, the Facebook app's page as shown in \myfig~\ref{fig:playPage}.
We can notice that three attributes (``Size'', ``Current Version'', and ``Requires Android'') all have the same value of ``Varies with device''. 
This indicates that Facebook employs the multiple-apk approach to handle the app compatibility over different versions of Android platforms.
The apps that do not have the value of ``Varies with device'' are thus the single-apk apps.
%
%
%
%

To implement the Google Play analysis, we write Python scripts based on our previous codes~\cite{FileCross14,MoST15} and Selenium, a web browser automation tool.
We use Selenium's Firefox driver to load each app's Google Play page, and extract the attribute values we are interested by parsing the page's HTML source.


\noindent
\textbf{Dataset}
Table \ref{tab:dataset} lists the dataset used in this paper.
We have crawled 24,426 apps from Google Play in July 2015.
We run Google Play analysis for all these apps, among which we identify and filter 1,301 multiple-apk apps.
Therefore, the remaining 23,125 single-apk apps assemble our final dataset, which will be further analyzed in Section \ref{sec:AppAnalysis}.
Unless stated otherwise, we refer to our dataset as these 23,125 apps in this paper.



\subsection{Android Document Analysis}
\label{sec:DocAnalysis}


\noindent
\textbf{Method}
To build the API-SDK mapping, we analyze Android SDK documents based on a previous work~\cite{ICSM13}.
Specifically, we first build a list of all Android APIs and the corresponding platform versions they were introduced to by parsing a SDK document called \texttt{api-versions.xml}.
This file covers both initial APIs (those introduced in the first Android version) and other newly added APIs in subsequent Android versions.
We further count the API change (e.g., deprecated and removed APIs) by analyzing the HTML files in the \texttt{api\_diff} directory.

After running the document analysis for 23 Android versions (from 1.0 to 6.0), we recorded a total of 30,083 APIs, out of which 794 APIs were afterwards deprecated and 190 APIs were finally removed.
However, we found that the lists of deprecated and removed APIs are not fully accurate, probably due to the mistakes made by Google developers when they wrote SDK documents.
For example, the \texttt{removeAccount(Account, Callback, Handler)} API in the \texttt{AccountManager} class was recorded as ``removed in SDK version 22'' in the documents, but actually it is still available in the SDK version 23.
This result implies that such a document-based analysis employed by the previous work~\cite{ICSM13} requires further improvement.
As a future work, we will explore to retrieve the API-SDK mapping directly from each SDK \texttt{jar} file.
In this paper, since the list of added APIs is accurate, we use only this part of results for the subsequent \DSDK analysis in Section~\ref{sec:evaluate}.

\begin{figure}[t!]
\begin{adjustbox}{center}
    \begin{minipage}{0.35\textwidth}
        \centering
        \includegraphics[width=1\textwidth]{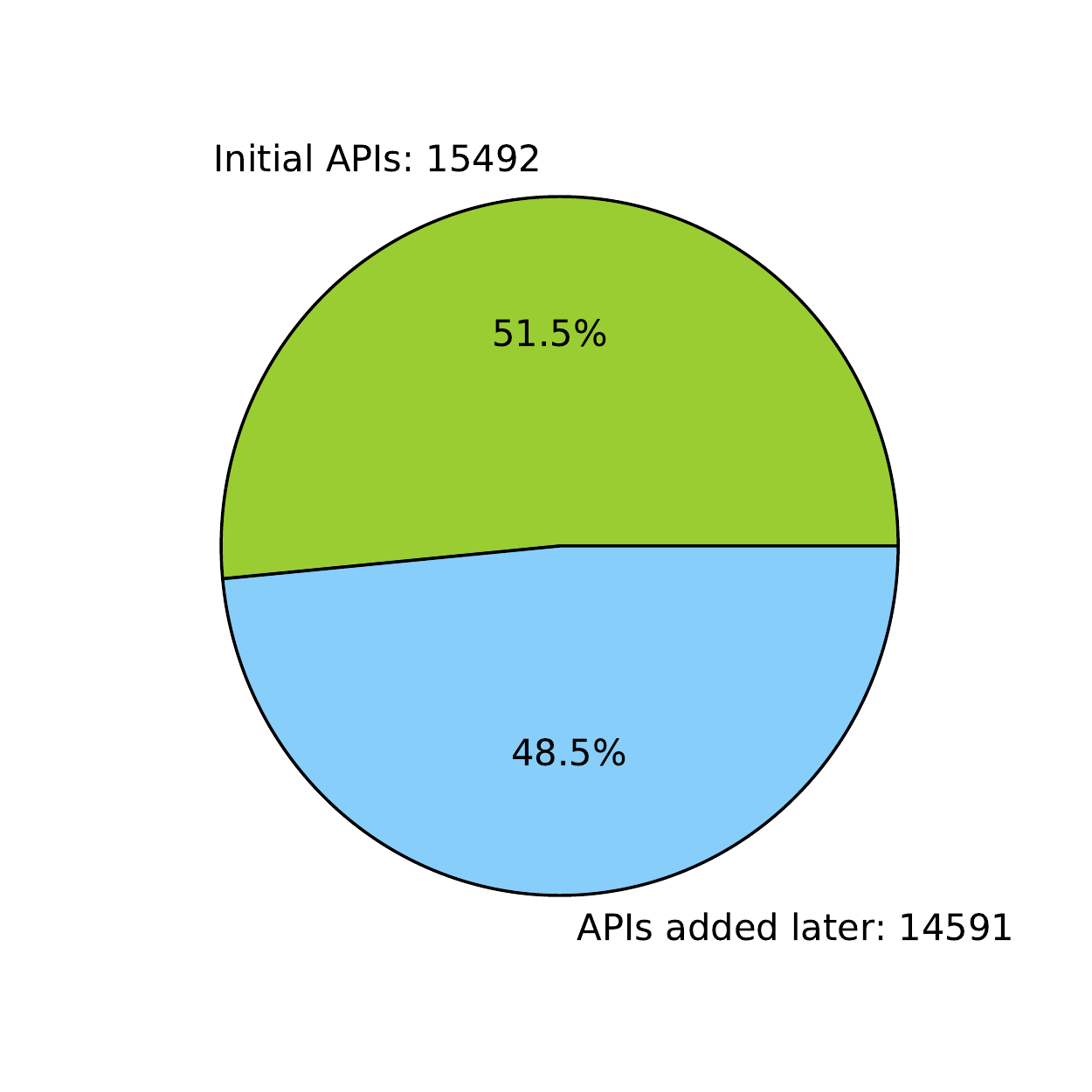}
        \caption{\footnotesize The comparison between initial and added APIs.}
        \label{fig:APIadd}
    \end{minipage}
    \begin{minipage}{0.65\textwidth}
        \centering
        \includegraphics[width=1\textwidth]{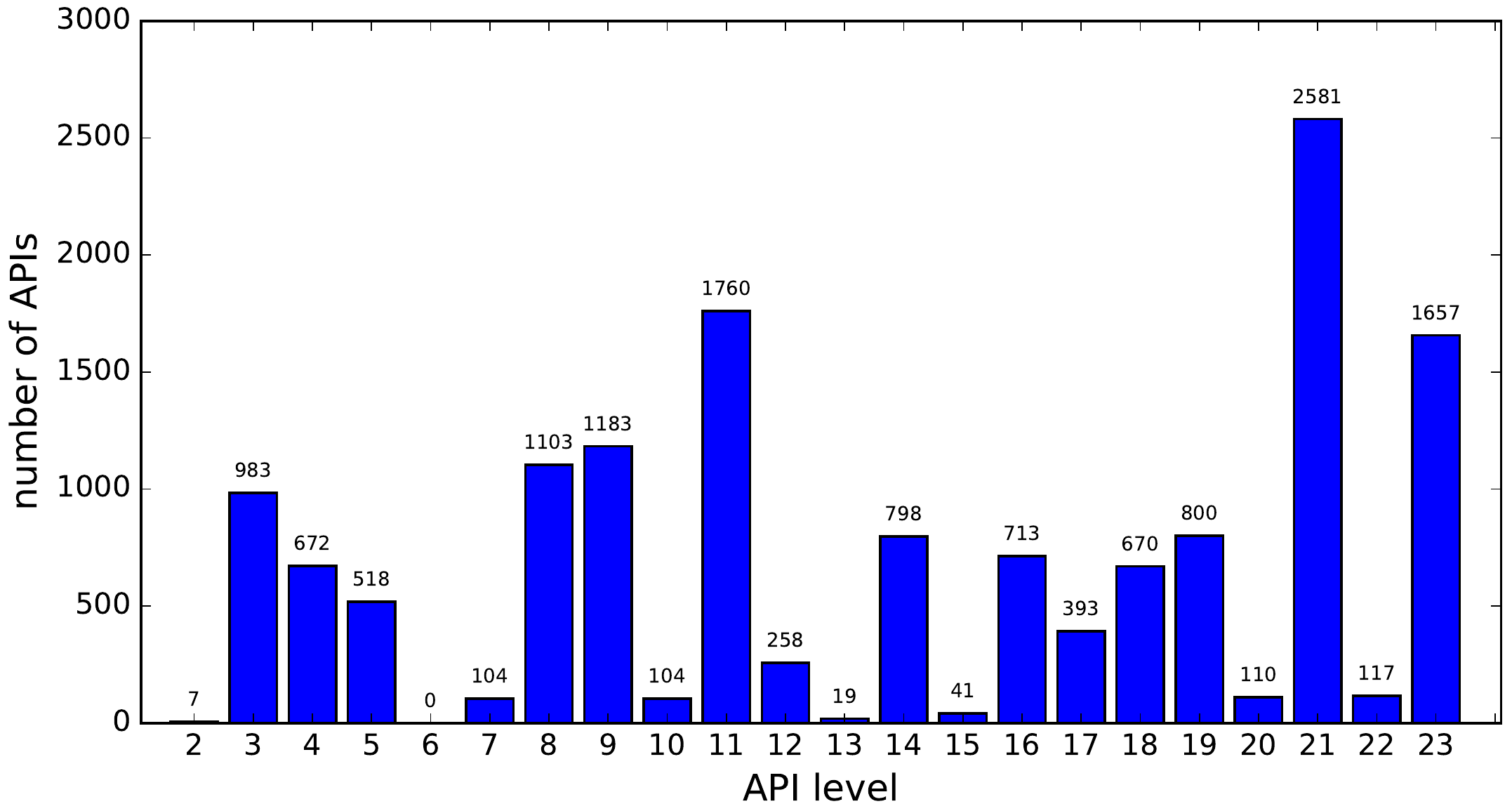}
        \caption{The distribution of added Android APIs.}
        \label{fig:APIlevel}
    \end{minipage}
\end{adjustbox}
\end{figure}

\noindent
\textbf{Results}
We now present the results of document analysis.
\myfig~\ref{fig:APIadd} shows the comparison between the initial Android APIs and those subsequently added APIs.
We can see that almost half of all APIs were added afterwards.
This indicates that Android evolves dramatically along the whole process. 
In \myfig~\ref{fig:APIlevel}, we further plot the distribution of those subsequently added APIs since API level~2.
Android 5.0 (API level 21) changed most, with 2,581 new API introduced.
The following two most changed versions are Android 3.0 (API level 11) and Android 6.0 (API level 23), with 1,760 and 1,657 new APIs, respectively.

\subsection{Android App Analysis}
\label{sec:AppAnalysis}


\noindent
\textbf{Retrieving Declared SDK Versions}
We leverage \aapt (Android Asset Packaging Tool) to retrieve \DSDK versions \textit{directly} from each app without extracting the manifest file.
This method is more robust than the traditional \apktool-based manifest extraction employed in many other works.
Indeed, our \aapt-based approach can successfully analyze all 23,125 apps, whereas a recent work \cite{ECVDetector14} shows that \apktool fails six times in the analysis of top 1K apps.

%
%

In the course of implementation, we observed and handled two kinds of special cases.
First, some apps define \minSDK multiple times, for which we only extract the first value.
Second, we apply the by-default rules (see Section \ref{sec:declared}) for the non-defined \minSDK and \aimSDK.
More specifically, we set the value of \minSDK to 1 if it is not defined, and set the value of \aimSDK (if it is not defined) using the \minSDK value.



\noindent
\textbf{Extracting API Calls and Their SDK Versions}
To extract API calls from apps' bytecodes, we first translate the compressed bytecodes into readable texts by using the \dexdump tool.
We then use a set of Linux bash commands to extract each app's method calls from their \dexdump outputs.

With the extracted API calls, we use the API-SDK mapping to compute their corresponding SDK versions (i.e., \minLevel and \maxLevel, as explained in \myfig~\ref{fig:demystify}).
To compute the \minLevel, we calculate a maximum value of all API calls' added SDK versions.
Similarly, to compute the \maxLevel, we calculate a minimum value of all API calls' removed SDK versions.
If an API is never removed, we set its removed SDK version to a large flag value (e.g., 100,000).

During the experiments, we find that it is necessary to exclude library codes' API calls from host apps' own API calls.
Libraries such as Android Support Library provide the stub implementation of higher-version APIs on lower-version platforms to ensure the backward-compatibility of higher-version APIs.
If an app is running on a higher-version platform, the library directly calls the corresponding API.
Otherwise, the library calls the stub implementation, which actually does nothing but would not crash the app.
Since we currently do not differentiate such control-flow information, we exclude library codes for the consistency analysis.


\noindent
\textbf{Comparing Consistency}
With the \DSDK and API level information, it is easy to compare their consistency.
We compute the following three kinds of inconsistency (as previously mentioned in Section~\ref{sec:sideeffect}):
\begin{itemize}

\item $\minSDK < \minLevel$: the \minSDK is set too low and the app would crash when it runs on platform versions between \minSDK and \minLevel.

\item $\aimSDK < \maxLevel$: the \aimSDK is set too low and the app could be updated to the version of \maxLevel. If the \maxLevel is infinite, the \aimSDK could be adjusted to the latest Android version.

\item $\maxSDK > \maxLevel$: the \maxSDK is set too large and the app would crash when it runs on platform versions between \maxLevel and \maxSDK.

\end{itemize}

\section{Evaluation}
\label{sec:evaluate}
\vspace{-2ex}

Our evaluation aims to answer the following three research questions:
\begin{description}
  \item[\textbf{RQ1}] What are the \textit{characteristics} of the \DSDK versions in real-world apps?

  \item[\textbf{RQ2}] What are the \textit{characteristics} of the API calls in real-world apps?

  \item[\textbf{RQ3}] Could we identify the \textit{inconsistency} between \DSDK versions and API calls in real apps?
    In particular, could we discover crash bugs and potential security vulnerabilities?
\end{description}

\vspace{-3ex}
\subsection{RQ1: Characteristics of the Declared SDK Versions}
\label{sec:rq1}

In this section, we report a total of four findings regarding the RQ1.

\textbf{Finding 1: Not all apps define the \minSDK and \aimSDK attributes, and 16.5\% apps do not claim the \aimSDK attributes.}
From Table \ref{tab:nondefined}, we can see that rare apps (about 0.22\%) do not define the \minSDK, while a noticeable portion of apps (over 15\%) do not define the \aimSDK.
Out of these apps, 48 apps declare neither the \minSDK, nor the \aimSDK.
Consequently, the values of both \minSDK and \aimSDK will be assigned to ``1'' by the system.
We also notice that almost all apps (over 99\%) do not define the \maxSDK.
This result is reasonable because, as we described in Section \ref{sec:declared}, the \maxSDK attribute is strongly suggested \textit{not} to define.

\begin{table}[t!]
\vspace{-2ex}
\caption{\small The number and percentage of non-defined \DSDK attributes in our dataset.}
\begin{adjustbox}{center}
\begin{tabular}{  c| c | c}

\hline
 & \# Non-defined & \% Non-defined \tabularnewline
\hline
\hline

\minSDK    & 51         & 0.22\% \tabularnewline
\hline
\aimSDK    & 3,826      & 16.54\% \tabularnewline
\hline
\maxSDK    & 23,109     & 99.93\% \tabularnewline
\hline

\end{tabular}
\end{adjustbox}
\label{tab:nondefined}
\vspace{-4ex}
\end{table}

\textbf{Finding 2: There are 53 outlier \aimSDK values.}
We also find out some declared \aimSDK are outlier values.
One app defines its \aimSDK as 0, which is lower than the \minSDK.
Others' \aimSDK are larger than the newest SDK version (API level 23 at that time).
Some apps declare \aimSDK as 24, 25, 26 or larger, however, these SDK versions have not been released yet in year 2015.
Even more surprisingly, one app sets the \aimSDK value to ``10000''.
In general, \aimSDK should be always greater than or equal to the \minSDK, but 34 apps have negative \aimSDK - \minSDK value.

\begin{figure}[t!]
\vspace{-2ex}
\begin{adjustbox}{center}
    \begin{minipage}{0.53\textwidth}
        \centering
        \includegraphics[width=1\textwidth]{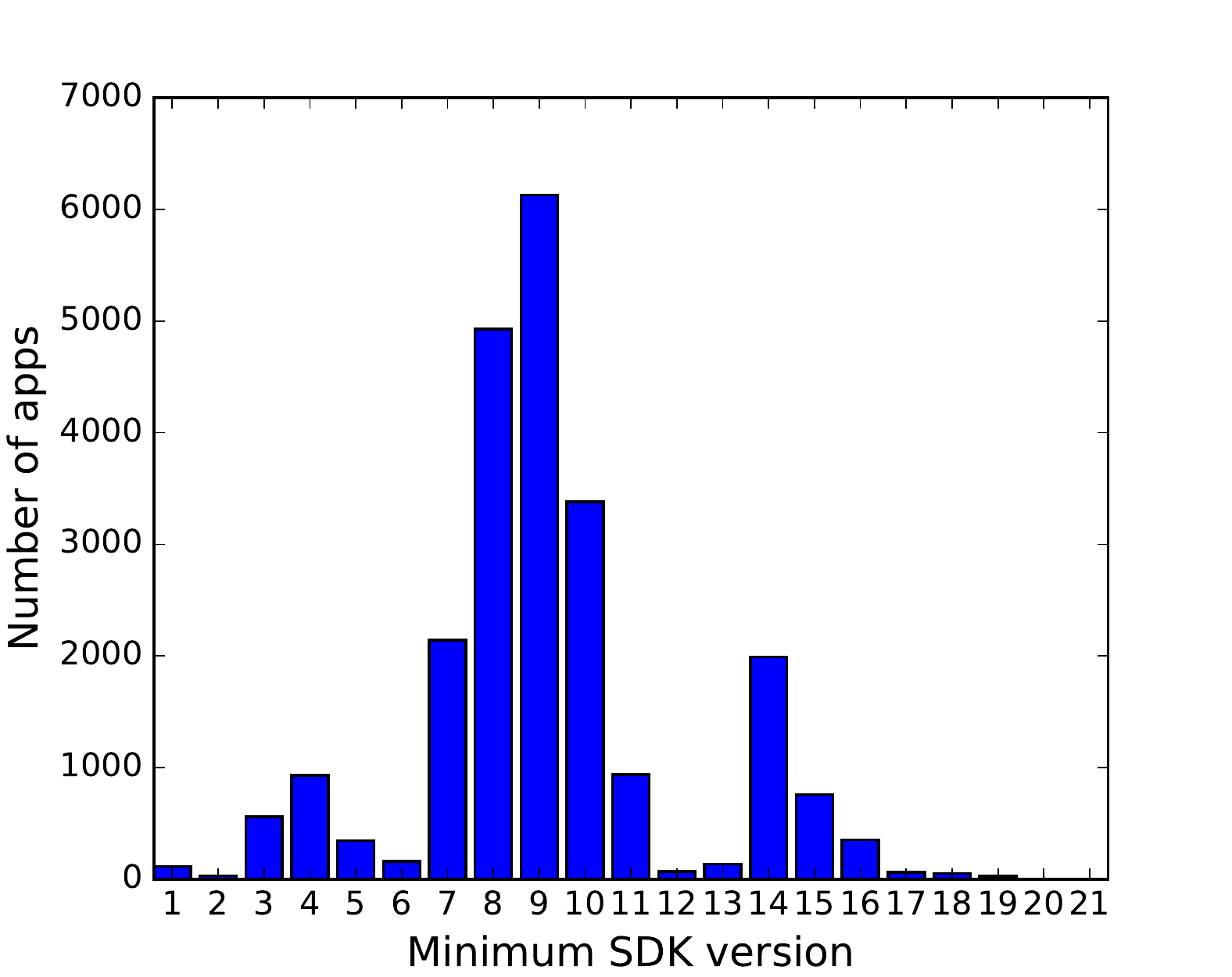}
        \vspace{-4ex}
        \caption{Distribution of \minSDK.}
        \label{fig:minSDKdestribute}
    \end{minipage}
    \begin{minipage}{0.53\textwidth}
        \centering
        \includegraphics[width=1\textwidth]{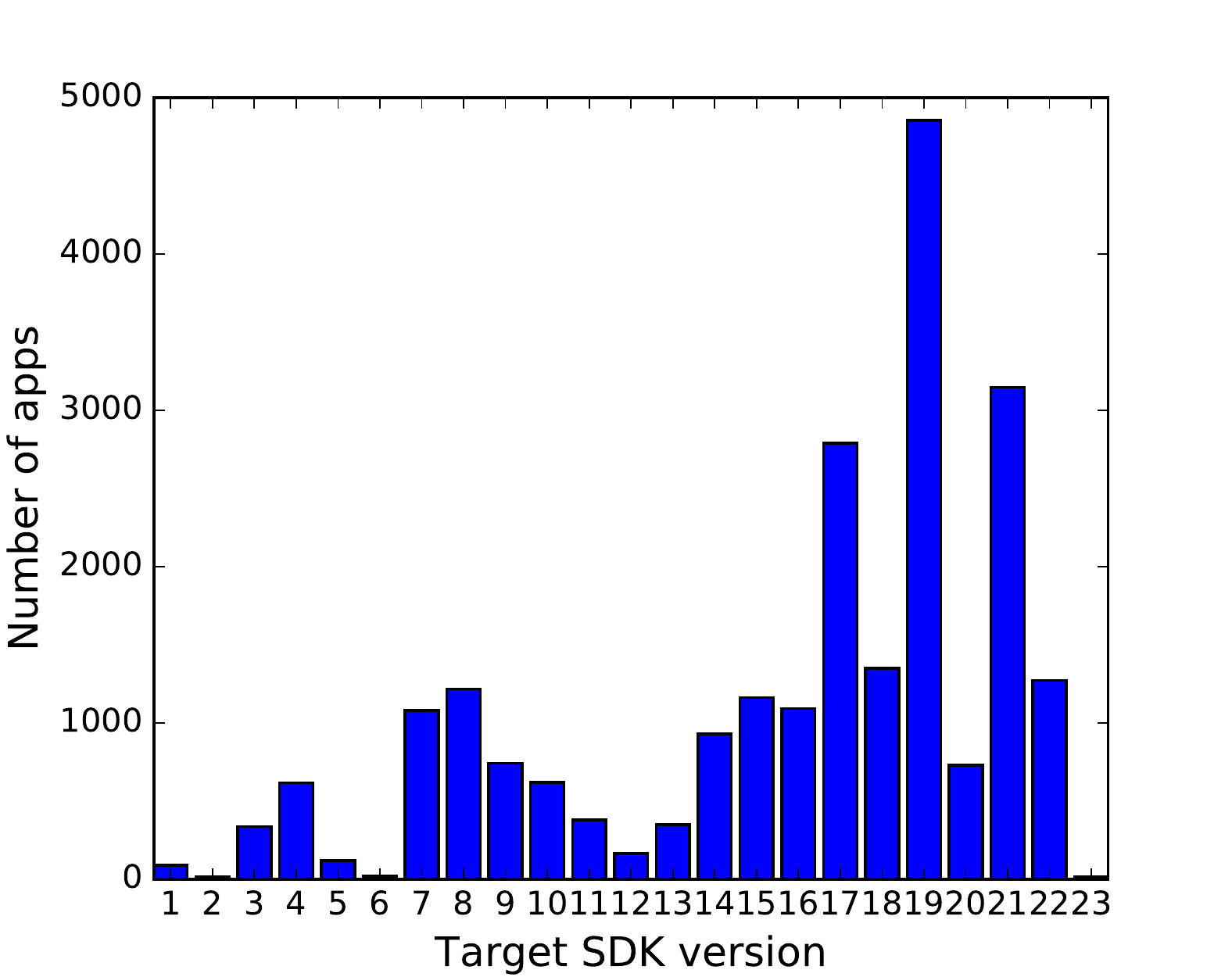}
        \vspace{-4ex}
        \caption{Distribution of \aimSDK.}
        \label{fig:tarSDKdestribute}
    \end{minipage}
\end{adjustbox}
\vspace{-2ex}
\end{figure}

\textbf{Finding 3: The minimal platform versions most apps support are Android 2.3 and 2.2, whereas the most targeted platform versions are Android 4.4 and 5.0.}
In \myfig \ref{fig:minSDKdestribute} and \myfig \ref{fig:tarSDKdestribute}, we plot the distribution of \minSDK and \aimSDK, respectively.
We can see that most apps (around 85\%) have \minSDK lower than or equal to level 11 (i.e., Android 3.0), which means that they can run on the majority of Android devices in the market~\cite{dashboards}.
Moreover, the minimal platform versions most apps support are Android 2.3 and 2.2.
\myfig \ref{fig:tarSDKdestribute} shows that more than 89\% apps test their apps on platform versions larger than Android 4.0, and the most targeted platform versions are Android 4.4 and 5.0.

\begin{figure}[t!]
\begin{adjustbox}{center}
    \begin{minipage}{0.36\textwidth}
        \centering
        \includegraphics[width=1\textwidth]{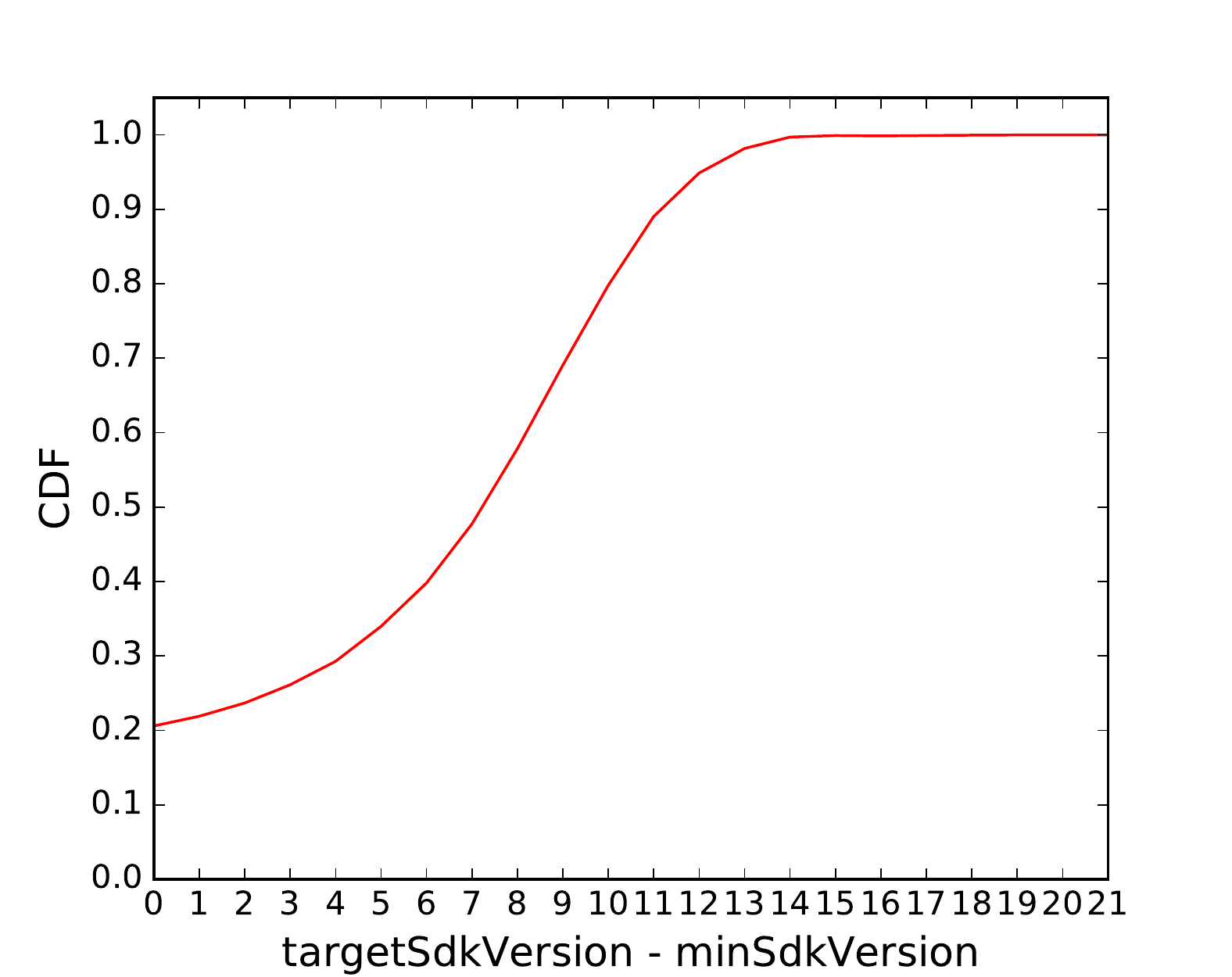}
        \caption{\small CDF plot of \lagSDK.}
        \label{fig:lagSDK}
    \end{minipage}
    \begin{minipage}{0.46\textwidth}
        \centering
        \includegraphics[width=1\textwidth]{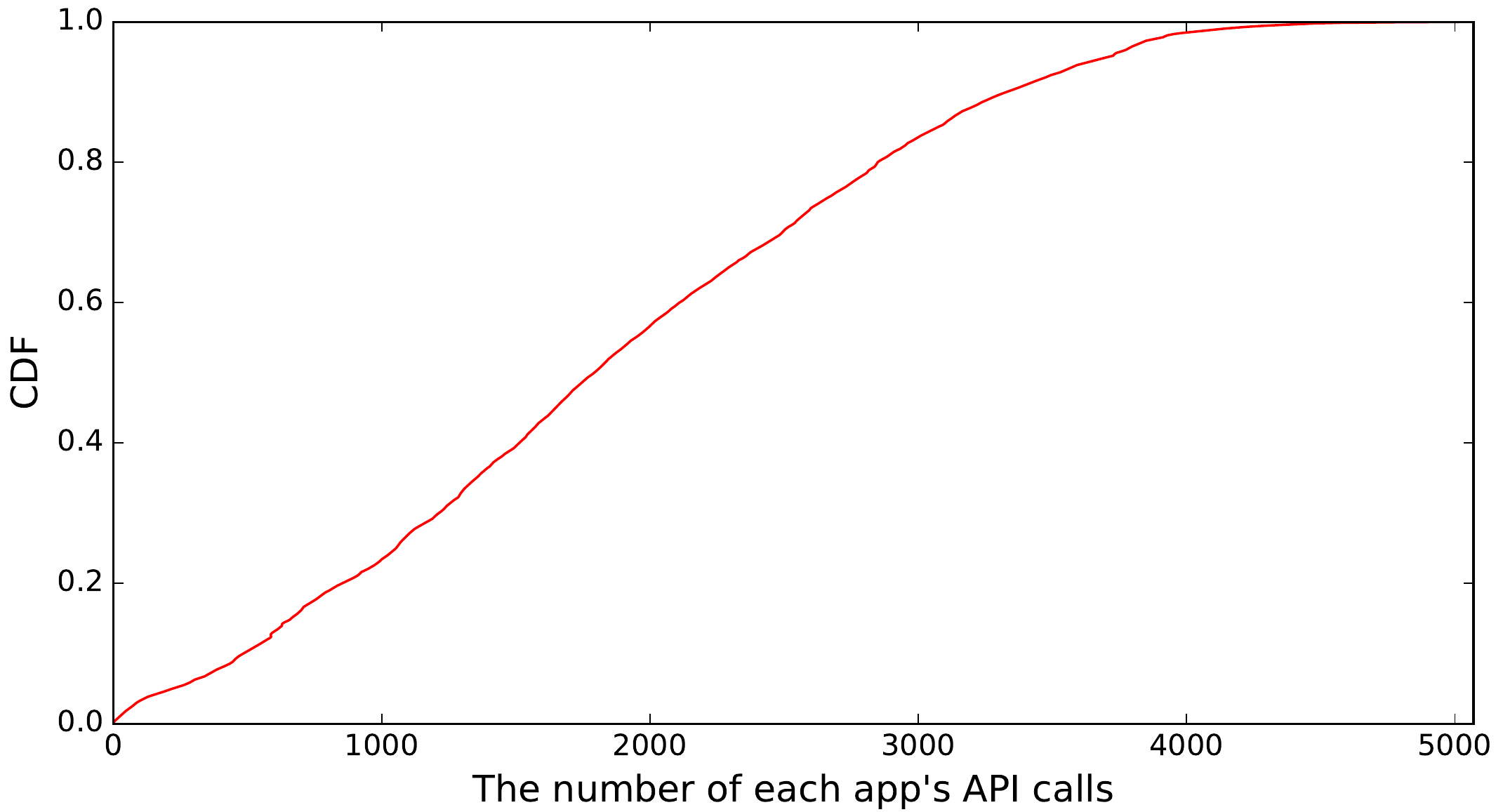}
        \caption{\small CDF plot of the number of each app's API calls.}
        \label{fig:APInum}
    \end{minipage}
    \begin{minipage}{0.36\textwidth}
        \centering
        \includegraphics[width=1\textwidth]{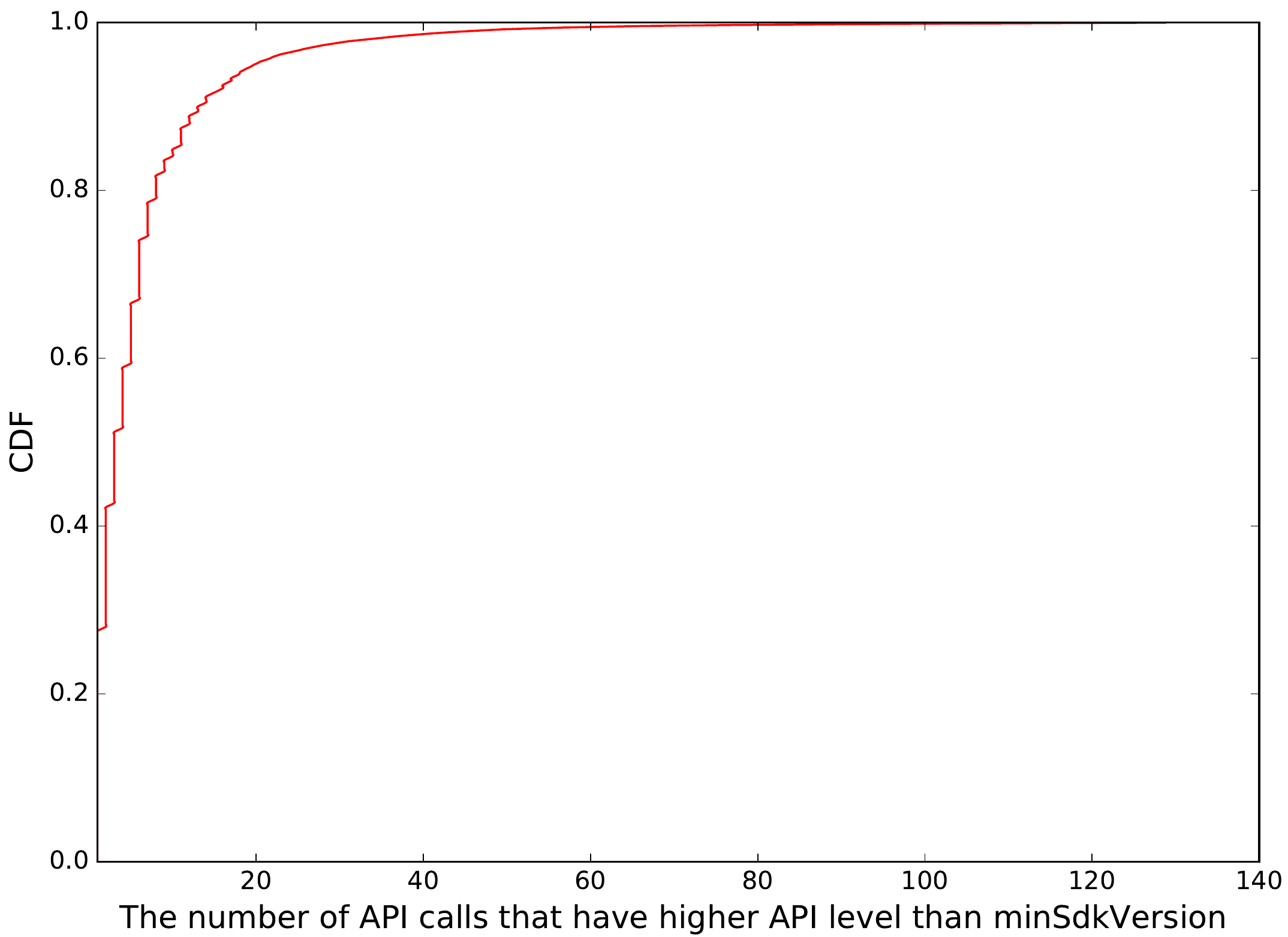}
        \caption{\small CDF plot of each app's number of API calls that have higher API level than \minSDK.}
        \label{fig:minOverNum}
    \end{minipage}
\end{adjustbox}
\vspace{-2ex}
\end{figure}

\textbf{Finding 4: The mean version difference between \aimSDK and \minSDK is 8.}
We define a new metric called \lagSDK to measure the version difference between \aimSDK and \minSDK, as shown in Equation \ref{equ:lagversion}.
\begin{equation}
  \lagSDK = \aimSDK - \minSDK
\label{equ:lagversion}
\end{equation}
After removing negative \aimSDK values and outliers, we draw the CDF (Cumulative Distribution Function) plot of \lagSDK in \myfig \ref{fig:lagSDK}.
It shows that more than 20\% apps have equal \aimSDK and \minSDK.
Furthermore, the majority of apps (more than 95\% apps) have a \lagSDK less than 12.

\subsection{RQ2: Characteristics of the API Calls}
\label{sec:rq2}

In this section, we briefly present two more findings related to the RQ2.
It is worth noting that here we consider all API calls that include the API calls in libraries.

\textbf{Finding 5: Around 500 apps call less than 50 APIs, making them lightweight apps. On the other hand, half of apps call over 1.8K APIs.}
We find that 446 apps call less than 50 APIs.
The majority of them are about user interface improvement, such as system theme and wallpaper apps.
These apps are regarded as lightweight ones that have less dependency on the SDK versions.
Additionally, many other apps contain several thousand API calls.
We plot the distribution of apps by API call numbers in \myfig~\ref{fig:APInum}.

\begin{figure}[t!]
\vspace{-2ex}
\begin{adjustbox}{center}
  \subfigure[All API calls with library code.] {
	\label{fig:allMinimumAPIlevel}
    \includegraphics[height = 33ex]{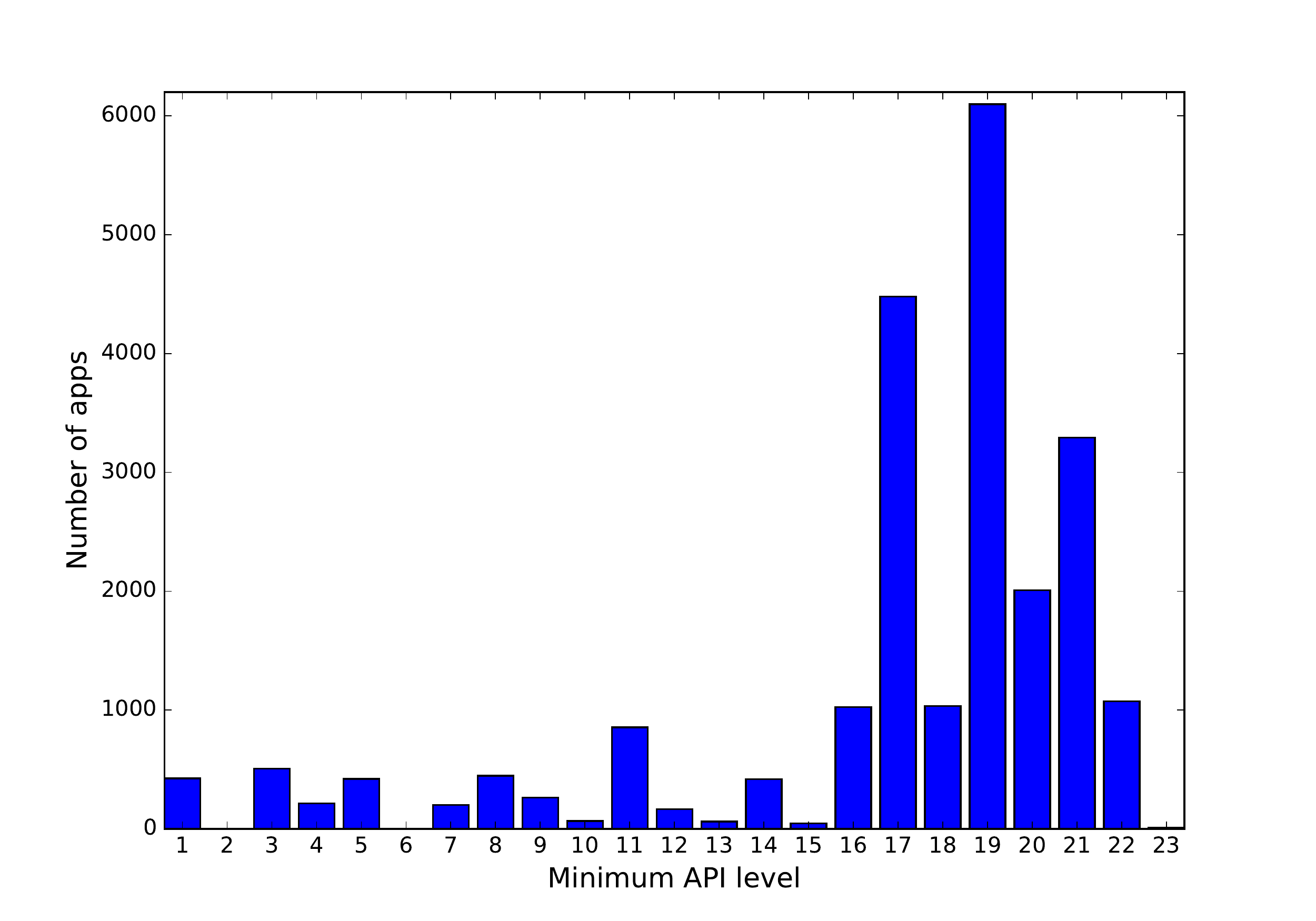}
  }
  \subfigure[App's own API calls without library code.] {
	\label{fig:ownMinimumAPIlevel}
    \includegraphics[height = 33ex]{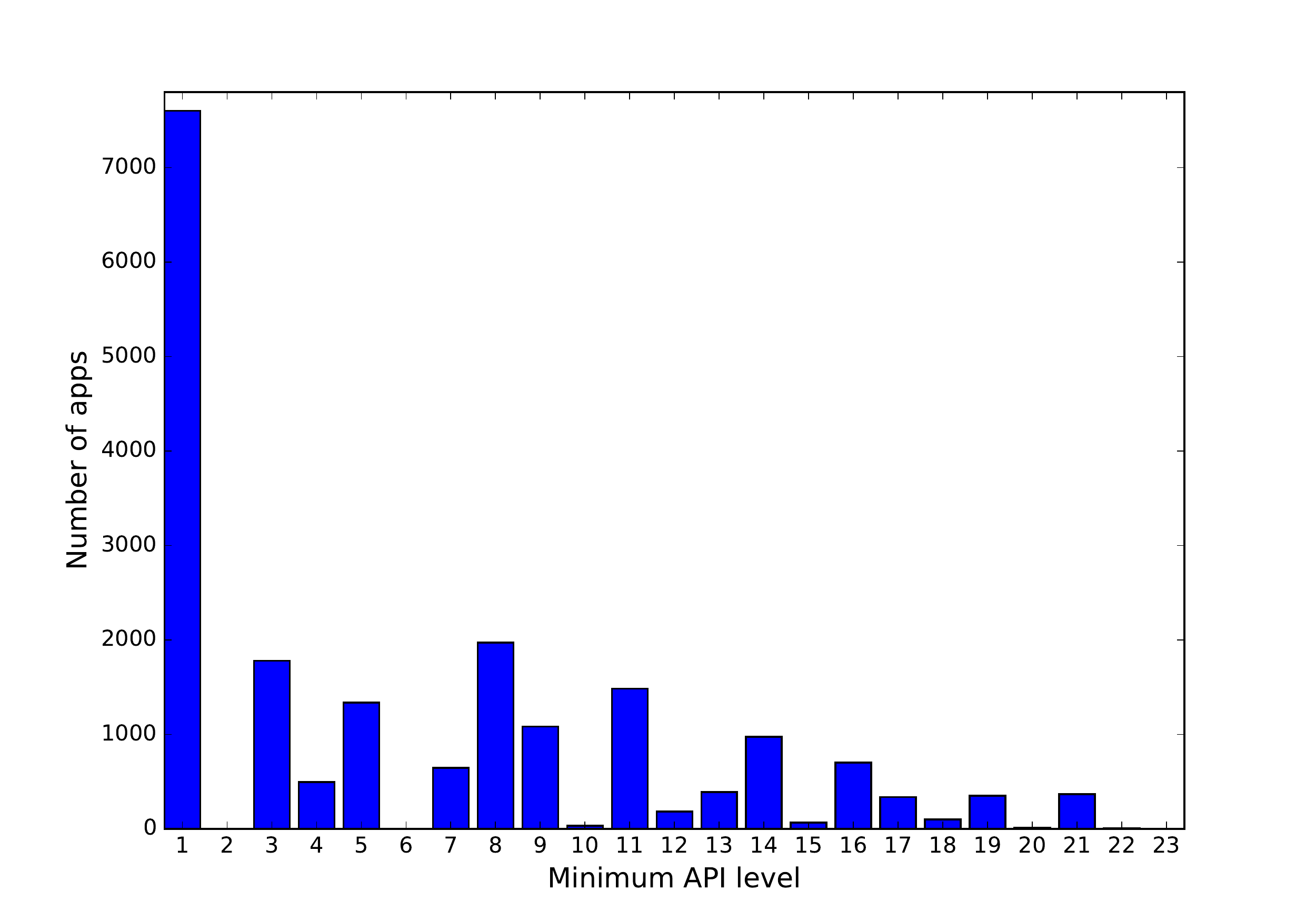}
  }
\end{adjustbox}
\vspace{-3ex}
\caption{\small The distribution of \minLevel that is calculated from API calls w/o library.}
\label{fig:MinimumAPIlevel}
\end{figure}

\textbf{Finding 6: Library codes contribute more higher-version API calls than apps' own codes.}
Libraries such as Android support library provide backward-compatible versions of Android framework APIs, as well as the features that are only available through the library APIs.
Each support library is backward-compatible to a specific API level, which allows an app that contains higher-version APIs run correctly on a lower version of Android system.
\myfig~\ref{fig:allMinimumAPIlevel} shows that distribution of the \minLevel of API calls with the library code, whereas \myfig~\ref{fig:ownMinimumAPIlevel} presents the distribution of the \minLevel of API calls without the library code. 
By analyzing and de-compiling the support library, we found that they can redirect the APIs calls in a higher-version SDK to some similar APIs which are already in a lower SDK or to an empty function.

\subsection{RQ3: Inconsistency Results}
\label{sec:rq3}

In this section, we report two important findings regarding the RQ3.

\textbf{Finding 7: Around 1.8K apps under-set the \minSDK value, causing them would crash when they run on lower Android versions.}
We find that 1,750 apps have over five API calls, the levels of which are larger than the declared \minSDK.
In 692 apps, more than ten API calls have higher API level than \minSDK.
In \myfig~\ref{fig:minOverNum}, we draw the CDF plot of the number of API calls that have higher API level than \minSDK.
Based on this figure, we find that several apps have more than 50 API calls whose API level is higher than \minSDK.


%
%

\textbf{Finding 8: Around 400 apps fail to update their \aimSDK values, making them potentially exploitable by remote code execution.}
The \texttt{addJavascriptInterface()} API \cite{addJavascriptInterfaceSaga} has a serious security issue.
By exploiting this API, attackers are able to inject malicious codes, which may obtain any information from SD card.
Google later fixed this bug on Android 4.2 and afterward.
However, as mentioned in the side effect II, if an app has the \aimSDK lower than 17 and calls this API, the system will still call the vulnerable API even when running in Android 4.2 and afterward.
In our dataset, we find that 909 apps call the \texttt{addJavascriptInterface()} API.
Among these apps, 413 apps are vulnerable, which may cause privacy information leakage.
In particular, out of these 413 apps, 238 apps do not define the \aimSDK attribute (i.e., \aimSDK is null).


\section{Threats To Validity}
\label{sec:discuss}

In this section, we discuss a couple of threats to the validity of our study.

First, we have not performed the control-flow analysis to determine whether an API call will be invoked only when running on certain Android versions.
During the experiments, we noticed that many library codes take \texttt{if-else} blocks to call higher-version APIs on when the app is running on the corresponding versions.
To mitigate its impact to our analysis, we currently exclude the library codes for consistency analysis (Section~\ref{sec:AppAnalysis}), and use a threshold value to minimize the potential version-related \texttt{if-else} blocks in app codes (Section~\ref{sec:rq3}).

Apps may employ Java reflection to call private Android APIs \cite{privateAPI} that are not included in the SDK but contained in Android framework. 
Similarly, developers may use native codes to access Android APIs.
Currently we have not handled these two cases and leave them as our future work.

Our assumption in Section~\ref{sec:overview} that multiple-apk apps do not have compatibility issues may not be always true.
In particular, developers may provide only one apk for several Android platforms to share. 
In this case, those shared apks are similar to single-apk apps.

\section{Related Work}
\label{sec:related}

Our paper is mainly related to prior works that also study Android APIs or SDKs.
The work performed by McDonnell et al.~\cite{ICSM13} is the closest to our paper.
They studied the Android API evolution and how client apps follow Android API changes, which is different from our focus on the consistency between apps' \DSDK and API calls.
In the methodology part, we followed their document analysis method for extracting the API-SDK mapping.
But in the future we plan to directly analyze Android SDKs instead of documents for more accurate mapping extraction.
Other related works have studied the coefficient between apps' API change and their success~\cite{AppSuccess13}, the deprecated API usage in Java-based systems~\cite{DeprecatedAPI16}, and the inaccessible APIs in Android framework and their usage in third-party apps~\cite{Inaccessible16}.
Two recent works~\cite{TargetFragment16,TameFragment16} also focused on the fragmentation issues in Android.
Compared to all these works, our study is the first systematic work on \DSDK versions and their consistency with API calls.


\section{Conclusion and Future Work}
\label{sec:conclude}
\vspace{-1.5ex}

In this paper, we made a first effort to systematically study the declared SDK versions in Android apps, a modern software mechanism that has received little attention. 
We measured the current practice of the declared SDK versions or \DSDK versions in a large dataset of apps, and the consistency between the \DSDK versions and their app API calls.
To facilitate the analysis, we proposed a three-dimensional analysis method that operates at both Google Play, Android document, and Android app levels.
We have obtained some interesting and novel findings, including (i) around 17\% apps do not claim the targeted \DSDK versions or declare them wrongly, (ii) around 1.8K apps under-set the minimum \DSDK versions, causing them would crash when running on lower Android versions, and (iii) over 400 apps under-claim the targeted \DSDK versions, making them potentially exploitable by remote code execution.
In the future, we plan to contact the authors of the apps to inform them about the detected issues and collect their feedback, release a publicly available tool to let app developers detect and fix issues, and improve our approach to further mitigate the threats to validity (e.g., by designing and incorporating a suitable control-flow analysis technique).

\vspace{-1.5ex}

\begin{footnotesize}
\bibliographystyle{splncsnat}
\bibliography{main}
\end{footnotesize}

\end{document}